\documentstyle[aps,prl]{revtex}
\begin{document}
\input epsf.sty
\twocolumn[\hsize\textwidth\columnwidth\hsize\csname %
@twocolumnfalse\endcsname
\draft
\widetext
\title{Correlations and N\'eel Order of Randomly Diluted Quantum Spin Ladders}
\author{M. Greven$^{1,2}$ and R.J. Birgeneau$^2$}
\address{
$^1$ Department of Applied Physics and Stanford Synchrotron Radiation 
Laboratory, Stanford University, Stanford, CA 94305 \newline
$^2$ Department of Physics, Massachusetts Institute of Technology,
Cambridge, MA 02139}
\maketitle
\begin{abstract}
We present a Monte Carlo study of the correlation length
$\xi$ of randomly diluted
antiferromagnetic Heisenberg ladders, composed of two spin--1/2 chains.
For weak and intermediate inter--chain couplings, $J_{\bot}/J \leq 1$,
we find an enhancement of correlations that is strongest
for a fraction $z^* \approx J_{\bot}/(8J)$ of dilutants.
We are able to access the experimentally relevant low--temperature regime,
$T/J \approx 1/500$, and find that the
recently inferred N\'eel temperature
of $\rm Sr(Cu_{1-z}Zn_z)_2O_3$
corresponds to a curve of constant correlation length
$\xi \approx 18$
of the single diluted ladder with $J_{\bot}/J \alt 1/2$.
The primary reason for the N\'eel ordering
is argued to be a strong enhancement of two--dimensional correlations due to
a Cu--Sr--Cu exchange coupling of $\approx 10 meV$
in the stacking direction of the ladders.
\end{abstract}
\phantom{.}
]
\narrowtext

Low--dimensional quantum Heisenberg antiferromagnets exhibit many
unusual collective properties owing to the increased importance of quantum
fluctuations when the dimensionality of the system is less than three, and to
the continuous $O(3)$ spin rotational symmetry of Heisenberg interactions.
The one--dimensional (1D) Heisenberg chain and its two--dimensional (2D)
analog, the square--lattice antiferromagnet, have been studied
extensively for many decades.
Arguably, the most prominent examples in this context are the
lamellar copper oxides,
which in their undoped states are
square--lattice Heisenberg antiferromagnets with predominant
nearest--neighbor (NN) interactions
and a spin value of $S=1/2$
\cite{square_lattice}.
The properties of the NN Heisenberg chain are very different
from those of the square lattice.
Instead of true long--range order, the $S=1/2$ chain exhibits
a critical ground state characterized by
algebraic decay of the spin correlations and gapless excitations
\cite{Bethe}.
In 1983, Haldane \cite{Hal83}
conjectured that the basic properties of the $S=1/2$ chain are shared only by
chains with half--odd--integer spin, and that integer--$S$ chains
should behave fundamentally differently.
The latter were predicted to possess a spin--liquid ground state
characterized by a finite correlation length $\xi_0$
due to the presence of a gap $\Delta \sim 1/\xi_0$ to the lowest excitations.
These predictions have been confirmed both numerically and experimentally
\cite{Aff94}.

In recent years, the question of how quantum fluctuations manifest
themselves in systems whose extent is finite in one dimension and infinite in
another  has  received much attention.
In particular, arrays of coupled chains of width $n$,
so--called spin ladders, have been studied extensively \cite{review}.
Good physical realizations of these systems, with $S = 1/2$,
were found to exist in materials such as $\rm SrCu_2O_3$ ($n=2$) and
$\rm Sr_2Cu_3O_5$ ($n=3$) \cite{materials}.
In close analogy to the behavior of the spin--$S$ chains,
the properties of NN $S = 1/2$ ladders
are fundamentally different for even and odd numbers of coupled chains
\cite{review}.
Quite generally, it is believed
that quantum fluctuations always generate a spin gap when the
product $n S$ is an integer,
and that the destructive interference of fluctuations leads to a
gapless spectrum for half--odd--integer values of $n S$ \cite{Topology}.
Theory for the
temperature dependence of the spin correlations of $S=1/2$ ladders
\cite{Ladder2} is in good agreement with Monte Carlo results
\cite{Ladder2,Ladder1}.

A surprising recent experimental discovery has been that random dilution
in the ladder material $\rm Sr(Cu_{1-z}Zn_z)_2 O_3$
($S=1/2$, $n=2$) leads to
antiferromagnetic long--range order at low temperatures, 
even for very small Zn
concentrations \cite{Azuma}. A schematic of the structure of this material is
shown in Fig. 1(a). The observed N\'eel order 
is counter--intuitive, since
based on classical intuition one might have expected a diminution of
correlations due to dilution, and hence a reduced tendency to order.
A large number of numerical and theoretical studies
have addressed this issue \cite{Dilution}, yet
a quantitative understanding of the
experimentally relevant low--temperature
spin correlations of a diluted two--chain ladder is still lacking.
Furthermore, the explicit pathway by which dilution promotes three--dimensional
(3D) N\'eel order remains elusive.

In this paper, we present a numerical investigation of the effects of
random dilution $z$ on the spin correlations of a $n = 2,$ $\rm S=1/2,$ NN
antiferromagnetic Heisenberg ladder.
We are able to reach the experimentally relevant
low--temperature regime, $T/J \approx 1/500$,
and find that a small amount of dilution
indeed enhances the antiferromagnetic correlation length,
albeit rather weakly.
For weak and intermediate inter--chain couplings, $J_{\bot}/J \leq 1$, we find
that the enhancement is strongest
for an average spacing between dilutants that approximately equals
$\xi_0 = \xi(z=0,J_{\bot} / J, T/J = 0)$, 
the zero--temperature correlation length of the un--diluted
ladder. In the limit of small $J_{\bot} / J$, we have shown previously that
$\xi_0 \approx 3.9 J / J_{\bot}$ \cite{Ladder1}.  
Since the N\'eel temperature
$T_N(z)$  of $\rm Sr(Cu_{1-z}Zn_z)_2O_3$ peaks at 
$z \approx 4\%$ \cite{Azuma},
this in turn requires that $J_{\bot}/J \alt 1/2$.  
This bound is consistent with
recent experimental and theoretical estimates for the ratio $J_{\bot}/J$.
Moreover, we find that
$T_N(z)$ of
$\rm Sr(Cu_{1-z}Zn_z)_2O_3$ corresponds to a curve of constant 
correlation length $\xi \approx 18$
of a single ladder.
We argue that the primary reason for the experimentally observed
divergence of spin correlations at $T_N(z)$
is a crossover from 1D to 2D behavior due to
the inter--ladder Cu--Sr--Cu exchange
in the stacking direction of the ladders;
an important observation is the fact that the 
Cu--Sr--Cu geometry and distances are very similar to those in the
Cu--Y--Cu
bi--layer in the high--$T_c$ precursor
material $\rm YBa_2Cu_3O_{6+\delta}$.
From neutron scattering measurments (for $\delta = 0.15 - 0.20$)
it is known that the bi--layer coupling is $\approx 10 meV$ \cite{YBCO}. The
final 3D antiferromagnetic ordering then occurs because of a
very weak pseudo-dipolar coupling between the frustrated ladder segments
\cite{2342}.
We thus predict a broad 1D to 2D crossover of the spin
correlations followed by sharp 3D  N\'eel ordering transition.
This could be observed directly in neutron scattering experiments once 
suitably large single crystals become available.

As in our previous studies \cite{Ladder2,Ladder1},
the ladders are investigated with a very efficient loop cluster
algorithm \cite{MonteCarlo}.
The Hamiltonian operator for a Heisenberg ladder is
\begin{equation}
{\cal H} = J \sum_{\langle ij\rangle} {\bf S}_i \cdot {\bf S}_j
  + J_{\bot} \sum_{\langle ij\rangle_{\bot}} {\bf S}_i \cdot {\bf S}_j,
\label{H}
\end{equation}
where ${\bf S}_i = \frac{1}{2}{\bf\sigma}_i$ is the quantum spin operator
at each site $i$ of non--zero spin, while $\langle ij \rangle$ and
$\langle ij \rangle_{\bot}$ denote nearest neighbors along
the chains and across the rungs of the ladder,
respectively (see Fig. 1(b)).
The couplings considered are antiferromagnetic, that is, 
$J$ and $J_{\bot} > 0$.
Periodic boundary conditions are employed only along the chains, and
we use units in which the lattice constant is unity, as well as
$\hbar = k_B = g\mu_B = 1$.
We typically average over 40 random configurations of each
$10^4$ updates and measurements, and
compute the staggered instantaneous correlation function
\begin{equation}
C(i,j) = {\mbox{sign}(i,j)} \langle {\bf S}_i \cdot {\bf S}_j \rangle,
\end{equation}
where $\mbox{sign}(i,j) = 1$ if the spins at $i$ and $j$ are on
the same sublattice of the antiferromagnet, 
and $\mbox{sign}(i,j) = -1$ otherwise.
The correlation length of the diluted ladders,
$\xi = \xi(z,J_{\bot}/J,T/J)$,
is extracted at large distances $r$,
$r/\xi \geq 3$, from fits to the symmetrized form
$C(r) \sim \cosh[(L/2 - r)/\xi]$;
the length $L$ of the ladders is kept at least an order of magnitude
larger than $\xi$.

In the present study, we focus on ladders with
$1/4\leq J_{\bot}/J \leq1$.
The dependence of the correlation length on the degree of
dilution is shown in Fig. 2. Here, the temperature is chosen such that
$T/J_{\bot} = 3/100$ is constant.
As $z$ is increased from zero, $\xi$ is enhanced, reaches a maximum,
and eventually decreases at large $z$.
In a previous study of un--diluted ladders \cite{Ladder1}
we found that at weak inter--chain coupling the product $\Delta \xi$
exhibits a rather simple scaling behavior in the variable $T/\Delta$.
The weak--coupling regime
is characterized by the relationship $\Delta \approx 0.41 J_{\bot}$,
and it sets in below $J_{\bot}/J \approx 1/2$ \cite{NEW}.
This motivated us to re--plot the data of Fig. 2 as $\xi J_{\bot}/J$
versus $zJ/J_{\bot}$, as shown in the inset.
Indeed, the data for $J_{\bot} = 1/4$ and 1/3 collapse to
a single curve, while those for $J_{\bot} \geq 1/2$ deviate slightly from this
curve.

The enhancement of correlations appears to be largest for an approximate
average spacing between dilutants of
$\xi_0 = \xi(0, J_{\bot}/J,0)$; specifically
we find empirically that the maximum occurs for
$z^* = \frac{1}{2}\frac{1}{\xi_0 + \xi^*}$,
where $\xi^* = \cal{O}$(1) is a constant.
The weak--coupling regime is characterized by $\Delta \approx 0.41 J_{\bot}$
and $\xi_0 = c_{1D}/\Delta$, where $c_{1D}=\pi/2$ is the spin--wave
velocity of the single spin--1/2 chain \cite{Ladder1}.
Deep in the weak--coupling regime, $\xi_0 \approx
3.9J/J_{\bot}$  is large, so that $z^* \approx 0.13J_{\bot}/J$, consistent
with Fig. 2.

For $\rm Sr(Cu_{1-z}Zn_z)_2O_3$,
the N\'eel temperature, and thus the enhancement of correlations,
is the largest for $z \approx 4\%$.
Based on our observations for the single ladder, this implies that
$1/4 \leq J_{\bot}/J \leq 1/2$.
Such a strong exchange anisotropy might at first appear to be somewhat
surprising, given that the Cu--O--Cu distances along the chains and across
the rungs are nearly identical.  However, both recent experiments 
\cite{Joh96,Ecc97} and 
theoretical calculations \cite{theory} imply that $J_{\bot}/J$ is of order 
0.4 to 0.5,
consistent with our own deductions.

We now discuss quantitatively the N\'eel ordering of
$\rm Sr(Cu_{1-z}Zn_z)_2O_3$, based 
on the assumption that $J_{\bot}/J \approx 1/2$.
To effect this, we simulate the diluted
ladder with $J_{\bot}/J = 1/2$ down
to very low temperatures, $T/J \approx 1/500$.
In our previous study of the the un--diluted ladder ($z=0$) with 
$J_{\bot}/J = 1/2$ we had 
found that $\xi_0 \approx 7.5$ \cite{Ladder1}.
The temperature dependences of the correlation length 
at several non--zero values of $z$ are shown in Fig. 3 together with the 
previous result for $z=0$.
Corresponding staggered susceptibility data are given in the inset.  As noted
before, the enhancement of correlations is strongest for $z^* \approx 4\%$.

The maximum ordering temperature of the experimental system
is $T_N \approx 8K$ at $z^* \approx 4\%$ \cite{Azuma}, as shown in Fig. 4.
With $J \approx 1900K$ \cite{Joh96}, this corresponds to $T_N/J \approx 0.004$.
From Fig. 3, it is apparent that the correlations of the isolated ladder
for $z \approx 4\%$ and $T/J \approx 0.004$
are enhanced only to $\xi \approx 18$.
In fact, we find that $T_N (z)$ of $\rm Sr(Cu_zZn_{1-z})_2O_3$ corresponds to
a curve of constant correlation length, $\xi=18(1)$, as shown in Fig. 4.

What then is the mechanism that leads to the dramatic enhancement of
correlations and the concomitant N\'eel order observed experimentally?
The fact that the experimentally observed $T_N(z)$ corresponds
to a curve of constant correlation length of a single diluted ladder
implies that this mechanism is independent of the degree of dilution.
It is important to realize that
the inter--ladder Heisenberg coupling within the Cu--O
sheets remains essentially frustrated at non--zero dilution.
A key observation in understanding the onset of long--range order in
Zn--doped $\rm SrCu_2O_3$ is that the $\rm Cu^{2+}-Sr^{2+}-Cu^{2+}$
geometry (see Fig. 1(a))
and distances in the stacking direction of the ladders are
essentially identical to those in the $\rm Cu^{2+}-Y^{3+}-Cu^{2+}$
bi--layer arrangement in $\rm YBa_2Cu_3O_{6+\delta}$;
recent neutron scattering experiments, for
$\delta = 0.15-0.2$, give $J_c \approx 10meV$ \cite{YBCO}.
Therefore, the primary perturbation to the
quasi--1D ladders in $\rm Sr(Cu_{1-z}Zn_z)_2O_3$ is the  coupling between
ladders in adjacent planes.
The reduced value for this inter--ladder coupling
is $\alpha_{2D} = J_c/J
\approx 6\%$.
Accordingly, we anticipate a progressive crossover from 1D to 2D spin 
correlations, followed by a sharp transition to 3D N\'eel order due to a
intra--planar pseudo--dipolar coupling between adjacent ladders \cite{2342}.
This evolution should be observable experimentally once suitably large 
single crystals become available.

The inset of Fig. 3 shows some results for
the staggered susceptibility per spin,
$\chi_{s} \sim T^{-1} \langle (\sum_i  (-1)^i S_{i} ^{z} )^2 \rangle$.
We find that $J\chi_{s} \approx 160$ along the curve of constant
correlation length $\xi \approx 18$, Fig. 4.
Mean--field theory for the enhancement of the susceptibility predicts
\begin{equation}
\chi_{s}^{MF} = \frac{\chi_{s}}{1 - J_{c} \chi_{s}},
\end{equation}
which diverges for
$J\chi_{s} = 1/\alpha_{2D} \approx 18$.
From the inset of Fig. 3 it is clear, that such a small value of $J\chi_{s}$
is already reached at rather high temperatures.
Perhaps not surprisingly, mean--field theory predicts
2D ordering, and, implicitly, 3D ordering, at a temperature which is about an 
order of magnitude higher 
than that observed experimentally 
for all values of $z$. Empirically, the condition for the
N\'eel ordering  appears to be
$\xi = 1/\alpha_{2D} \approx 18$, rather than $J\chi_{s} = 1/\alpha_{2D}$;
this remains to be understood theoretically.

A previous numerical study of the susceptibilities of the
diluted ladder \cite{Miy97} extended to temperatures as low as
$T/J = 1/200$ and used 
Eq. (3) together with $J_{\bot}/J = 1$, $J = 1000K$, and 
$\alpha_{3D} \approx 2 \%$, to explain qualitatively the phase diagam of
$\rm Sr(Cu_{1-z}Zn_z)_2O_3$. Our results are consistent with this work,
but give a physically more complete and precise description.

Recently, Fujiwara {\it et al.} \cite{Fuj98} 
presented an interesting NMR study of $^{65}$Cu linewidths in 
$\rm Sr(Cu_{1-z}Zn_z)_2O_3$ at small Zn concentrations, $z \leq 0.5\%$, and 
and for temperatures $T \geq 20K$. 
Our results imply 
that these data cannot be understood with the model suggested 
by the authors of Ref. \cite{Fuj98}, since the latter model requires ladder
correlation lengths one to two orders of magnitude larger than 
those which we compute.
Further work is required to understand these most interesting NMR results.

In summary, we have numerically determined the correlation length of randomly
diluted spin--1/2 two--chain Heisenberg ladders down to the
experimentally relevant low--temperature regime, $T/J \approx 1/500$.
At weak and intermediate inter--chain couplings, $J_{\bot}/J \leq1$,
we find an enhancement of correlations that is strongest for a fraction
$z^* \approx J_{\bot}/(8J)$ of dilutants.
The recently inferred N\'eel ordering temperature $T_N (z)$ of 
$\rm Sr(Cu_{1-z}Zn_z)_2O_3$ is found to correspond well to a curve
of constant correlation length $\xi \approx 18$ of the 
single diluted ladder with $J_{\bot}/J \alt 1/2$.
We argue that the primary reason for the experimentally observed 3D N\'eel
ordering is an enhancement of 2D correlations due to an antiferromagnetic 
coupling $\alpha_{2D} = J_c/J \approx 6\%$ in the stacking 
direction of the ladders. Empirically, the condition for N\'eel ordering thus
appears to be $\xi = 1/\alpha_{2D}$. The 1D to 2D crossover is 
then followed by a
sharp transition to 3D N\'eel order due to a tiny intra--planar 
pseudo--dipolar coupling between adjacent ladders. Once sizable single 
crystals become available, this predicted evolution of the 
correlations should be observable in neutron
scattering experiments. 

We would like to thank A. Aharony, H. Fukuyama, T. Imai, D.C. Johnston, 
Y.--J. Kim, P.A. Lee, T.M.
Rice, and U.--J. Wiese for stimulating discussions and correspondence.
This work was supported by the NSF under Grant
No. DMR 97-04532, the Joint Services Electronics Program, 
and the International Joint Research Program of NEDO
(New Energy Industrial Technology Development Organization, Japan).


\begin{figure}
\centerline{\epsfxsize=3.25in\epsfbox
{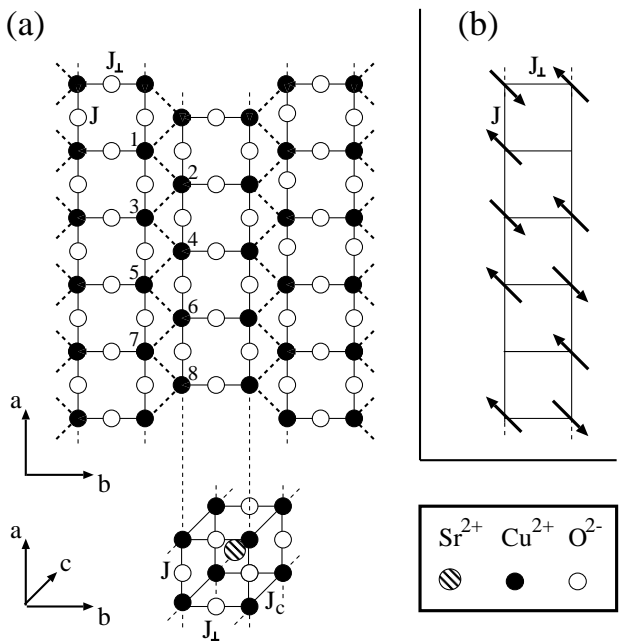}}
\vskip 1mm
\caption{
(a) Schematic structure of the two--chain ladder compound $\rm SrCu_2O_3$,
which has $\rm Cu^{2+}$ spin--1/2 magnetic moments
and antiferromagnetic exchange
couplings $J$ and $J_{\bot}$ along the chains 
and across the rungs, respectively.
The isotropic inter--ladder coupling, indicated 
by the dashed lines, is frustrated; at site 2, for example,
the coupling 1 - - 2 is cancelled by the coupling
2 - - 3.
$J_c$ is the Cu--Sr--Cu coupling in the stacking direction
of the ladders.
For $\rm Sr(Cu_{1-z}Zn_z)_2O_3$, a fraction $z$ of the
copper spins is randomly replaced by non--magnetic $\rm Zn^{2+}$ ions.
(b) Schematic of a single diluted ladder
as investigated in the present Monte Carlo study.
}
\label{ladders}
\end{figure}

\begin{figure}
\centerline{\epsfxsize=3.25in\epsfbox
{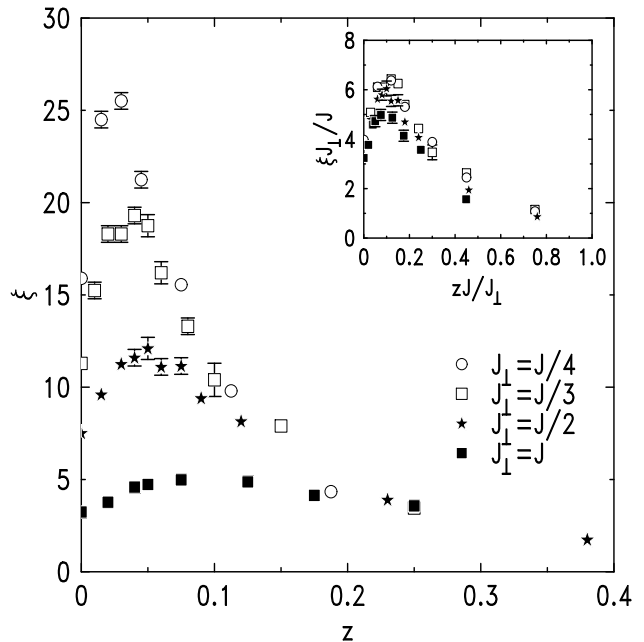}}
\vskip 1mm
\caption{
Correlation length versus degree of dilution of randomly diluted ladders.
The temperatures were chosen such that $T/J_{\bot}=3/100$.
Inset: Evidence for scaling of $\xi J_{\bot}/J$ in the variable
$zJ/J_{\bot}$.
}
\label{scaling}
\end{figure}

\begin{figure}
\centerline{\epsfxsize=3.25in\epsfbox
{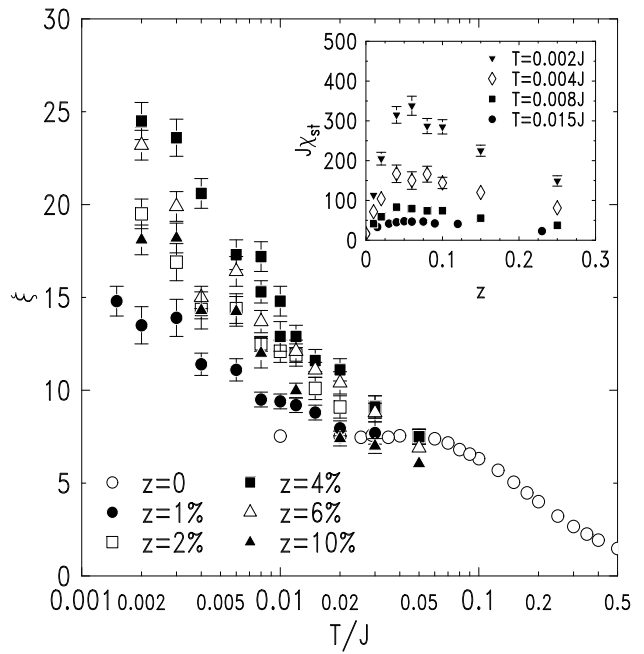}}
\vskip 1mm
\caption{Correlation length and staggered susceptibility (inset) for a
randomly diluted ladder with $J_{\bot}/J = 1/2$.
}
\label{xi}
\end{figure}

\begin{figure}
\centerline{\epsfxsize=3.25in\epsfbox
{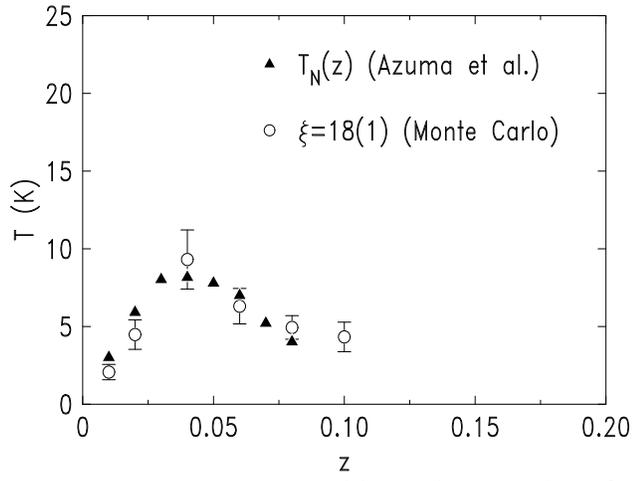}}
\vskip 1mm
\caption{N\'eel temperature $T_N (z)$
of $\rm Sr(Cu_{1-z}Zn_z)_2O_3$
\protect\cite{Azuma} and points of constant correlation length $\xi=18(1)$
of a single diluted ladder
with $J_{\bot}/J = 1/2$,
as obtained from Monte Carlo. In this comparison, we use $J = 1900K$
\protect\cite{Joh96}.
}
\label{T_N}
\end{figure}


\begin{references}
\bibitem{NEW}
In Ref. \protect\cite{Ladder1},
we obtained $\Delta = 0.41(1) J_{\bot}$;
based on additional data for $J_{\bot} = J/3$ and J/4 in the present study,
we find $\Delta = 0.407(5) J_{\bot}$.
\end{references}
\end{document}